\documentclass[aps,amssymb,amsmath,prl,superscriptaddress,reprint,longbibliography]{revtex4-2}
\usepackage{graphicx}
\usepackage{mathtools}
\usepackage[colorlinks,hyperindex]{hyperref}
	\hypersetup
	{
		colorlinks,%
		citecolor=black,%
		linkcolor=black,%
		urlcolor=black,%
	}

\newcommand{\affA}{London Centre for Nanotechnology, University College London, London, WC1H 0AH, UK}
\newcommand{\affB}{Department of Electronic \& Electrical Engineering, University College London, London, WC1E 7JE, UK}
\newcommand{\affC}{Quantum Motion Technologies, Nexus, Discovery Way, Leeds, West Yorkshire, LS2 3AA, UK}

\begin{document}
\title{Using deep learning to understand and mitigate the qubit noise environment}

\author{David F. Wise}\email{d.wise.15@ucl.ac.uk}\affiliation{\affA}\affiliation{\affC}
\author{John J. L. Morton}\affiliation{\affA}\affiliation{\affC}\affiliation{\affB}
\author{Siddharth Dhomkar}\email{s.dhomkar@ucl.ac.uk}\affiliation{\affA}

\date{\today}

\begin{abstract}
Understanding the spectrum of noise acting on a qubit can yield valuable information about its environment, and crucially underpins the optimization of dynamical decoupling protocols that can mitigate such noise. However, extracting accurate noise spectra from typical time-dynamics measurements on qubits is intractable using standard methods. 
Here, we propose to address this challenge using deep learning algorithms, leveraging the remarkable progress made in the field of image recognition, natural language processing, and more recently, structured data.
We demonstrate a neural network based methodology that allows for extraction of the noise spectrum associated with any qubit surrounded by an arbitrary bath, with significantly greater accuracy than the current methods of choice.
The technique requires only a two-pulse echo decay curve as input data and can further be extended either for constructing customized optimal dynamical decoupling protocols or for obtaining critical qubit attributes such as its proximity to the sample surface. Our results can be applied to a wide range of qubit platforms, and provide a framework for improving qubit performance with applications not only in quantum computing and nanoscale sensing but also in material characterization techniques such as magnetic resonance. 
\end{abstract}

\maketitle

Robust isolation of a qubit from unwanted noise in its environment is a key factor in technologies such as quantum computers and sensors. There is a long history in magnetic resonance of developing so-called dynamical decoupling protocols, including  Carr-Purcell-Meiboom-Gill (CPMG)~\cite{Hahn1950ab,Carr1954,Meiboom1958ab}, periodic dynamical decoupling (PDD)~\cite{Viola1998}, concatenated dynamical decoupling (CDD)~\cite{Khodjasteh2005}, and Uhrig dynamical decoupling (UDD)~\cite{Uhrig2007}, as general methods to preserve spin coherence in the presence of noise. In practice, the experimental noise spectrum varies significantly between different qubits in ways which are non-trivial to either predict, or indeed to accurately extract from most common measurements~\cite{Poggiali2018,Biercuk2009a,Alvarez2011}. 
As a result, it is difficult to predict \emph{a priori} which of the several possible dynamical decoupling protocols would provide optimal suppression of decoherence. Indeed, one could envisage constructing a decoupling protocol customized for a particular qubit, but this is impossible without knowing the actual qubit noise spectrum with sufficient accuracy. Additionally, it is worth emphasizing that the knowledge of obscured qubit environment is in itself a valuable outcome of a precise noise spectroscopy. 

Significant advances have been made in machine learning and specifically deep learning techniques, for example in the fields of computer vision and natural language processing, and more recently they have been applied to problems in physics and quantum engineering~\cite{Carleo2019}. Deep feed forward neural networks have been used to enhance extraction of material parameters in scanning probe microscopy~\cite{Borodinov2019}, for the derivation of the parameters associated with a $\frac{1}{f}$ noise spectrum from randomized benchmarking experiments~\cite{Zhang2019}, and for processing of magnetic resonance spectroscopy data, associated with nuclear magnetic resonance (NMR), electron paramagnetic resonance (EPR) and double electron-electron resonance (DEER) experiments~\cite{Qu2020, randbenchnn, Taguchi2019,Worswick2018,Youssry2020}. In quantum control, neural networks have been used to enhance dynamical decoupling protocols~\cite{optdyndec}, to suppress dephasing~\cite{mitigation}, and to improve the robustness of quantum control operations~\cite{robust}. In addition, deep reinforcement learning techniques have been applied to quantum metrology both as an efficient experiment design heuristic and to increase sensor sensitivity by more than order of magnitude over comparable approaches~\cite{Zhang2019,fiderer2020ab, Schuff2020}.

Here, we propose that the challenges of accurately obtaining qubit noise spectra can be efficiently handled by employing deep learning algorithms. We show how a deep neural network can be trained to extract the noise spectrum from simple and widely-used time-dynamics measurements on qubits, such as the two-pulse `Hahn' echo curves, and compare the accuracy of a deep learning approach with that of standard techniques. Finally, we examine a neural network based technique of processing noisy experimental data and discuss potential uses of an accurate noise spectrum for quantum control. In addition to the possibility of optimising dynamical decoupling to extend qubit coherence, we explore how useful information about the qubit environment such as its proximity to particular noise sources can be deduced from its noise spectrum~\cite{Romach2015}.

\section{Noise spectroscopy using dynamical decoupling}

\begin{figure*}[ht]
	\centering
	\includegraphics[width=1\linewidth]{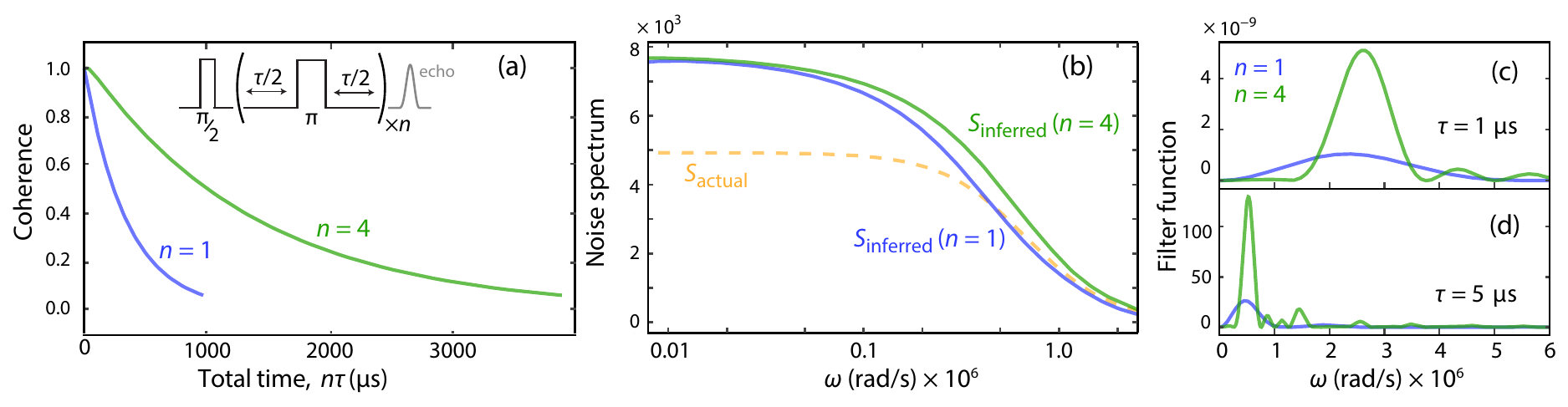}
	\caption{The challenge of extracting noise spectra from coherence decay measurements. (a) Decoherence curves under dynamical decoupling sequences (see inset) as a function of total sequence duration, $n\tau$, simulated using Eq.\ref{eq:2} from some underlying noise spectrum (dashed curve in (b)). Solid curves in (b) show noise spectra extracted from the coherence decays using the delta-function approximation to the filter function in Eq.\eqref{eq:3}. (c) and (d) Actual filter functions corresponding to sequences comprising one and four $\pi$-pulses at two $\tau$ values. Simulations assume a $\pi$-pulse duration of 100~ns.}
	\label{fig:Fig1}
\end{figure*}
The noise spectrum of a qubit is an effective proxy to probe its surroundings and provides valuable information for qubit characterization. Furthermore, once the environmental noise spectrum is known, it is possible in principle to run an optimization protocol that minimizes `decoherence' (which can be thought of as proportional to the overlap of the filter function $F(wt)$ associated with a given dynamical decoupling pulse sequence and the actual noise spectrum $S(\omega)$) by varying parameters in the sequence such as separation between pulses~\cite{Poggiali2018}, or their amplitude, phase, and duration. Alternatively, dynamical decoupling protocols can be tailored under real-time experimental feedback~\cite{Biercuk2009b}, however, such methods can be experimentally expensive to perform, limiting their applicability. Therefore, we focus here on using the \emph{minimum} experimental data from a given qubit, such as a single coherence decay curve under a particular pulse sequence, and using purely computational methods to identify the noise spectrum and optimized decoupling sequence.

The time dependence of qubit coherence under applied pulse sequences can be used for extracting useful information about the noise sources in the environment~\cite{Alvarez2011,Yuge2011,Bar-Gill2012,Hernandez-Gomez2018}. Such sequences, with overall duration of $t=n\tau$, can typically be described as a set of $n$ $\pi$-pulses, each with duration $\tau_\pi$ applied at time $t_k$, and possessing a characteristic filter function~\cite{Uys2009}:
\begin{equation} 
	\label{eq:1}
	F(\omega t) = \bigg{|}1+(-1)^{n+1}e^{i\omega n\tau}+2{\sum_{k=0}^{n} (-1)^{k}e^{i\omega t_{k}} cos\bigg{(}\frac{\omega \tau_{\pi}}{2}\bigg{)}}{\bigg{|}}^{2}
\end{equation}

Provided that the noise spectrum, $S(\omega)$, acting on the qubit is stationary, Gaussian, and couples only along the qubit's z-axis it can be extracted from the measured time-dependent coherence decay curve, C(t), by solving the following integral equation:

\begin{equation} 
	\label{eq:2}
	\chi(t) = - ln \: C(t) =  \int_{0}^{\infty}\frac{d\omega}{2\pi}S(\omega)\frac{F(\omega t)}{\omega^{2}}
\end{equation}
where, $\chi(t)$ is known as the decoherence functional~\cite{Cywinski2008}. However, in practice accurately solving such an integral equation is non-trivial,
and without a sufficiently faithful noise spectrum, the technique cannot yield an optimized protocol to significantly suppress decoherence. 

It is possible to simplify Eq.~\eqref{eq:2} by assuming the filter function at a given delay time to be a Dirac $\delta$-function localized at a desired frequency $\omega_0$ (which for CPMG sequences corresponds to $\tfrac{n\pi}{t}$) ~\cite{Bar-Gill2012}. This assumption permits a simple mapping of the coherence decay curve on to a corresponding noise spectrum:
\begin{equation} 
	\label{eq:3}
	S(\omega_0) = \frac{- \pi \: ln \: C(t)}{t}
\end{equation}
Figure~\ref{fig:Fig1} shows an application of this simplified approach to obtain noise spectra corresponding to decoherence curves that are simulated from some underlying noise spectrum, using the integral equation in Eq.~\ref{eq:2}. However, as demonstrated in Fig.~\ref{fig:Fig1}(c) the delta-function approximation is valid only where the spacing between pulses is much longer than the pulse duration. Increasing the number of  $\pi$-pulses produces a narrower peak in the filter function but with the expense of increased harmonics. As a result, the noise spectra inferred from the decoherence curves using Eq.~\ref{eq:3} are a poor fit to the actual spectrum used to generate them.   

A second challenge in obtaining accurate noise spectra from coherence decay curves is handling experimental noise in the measurement itself in an unprejudiced way. Depending on the type of qubit, the dominant noise sources can vary significantly, e.g.\ in the case of flux qubits and Si quantum dots the primary source of noise behaves as  ${1}/{f}^{\alpha}$ where $f$ is the frequency and $\alpha \sim 1$~\cite{Bylander2011,Yoneda2018}, telegraphic noise~\cite{Medford2012} dominates for GaAs quantum dots, bulk nitrogen vacancy (NV) centers in diamond are affected by Lorentzian-type noise~\cite{Klauder1962,Bar-Gill2012}, whereas near-surface NV centers are prone to double-Lorentzian-type noise~\cite{Romach2015}. Despite this rich variety of noise sources,  decoherence curves are often fit with a stretched exponential function with essentially two parameters – coherence lifetime, $T_2$, and power of the stretched exponential, $p$:
\begin{equation} 
\label{eq:4}
C(t) =  e^{-(t/T_2)^p}
\end{equation}
Whilst this equation holds for decoherence in the presence of certain simple noise forms (e.g. for simple $1/f^{\alpha}$ noise)~\cite{Medford2012}, in practice, qubits may be surrounded by multiple noise sources with varied functional forms, which can only be approximately represented by Eq.\eqref{eq:4}, and fitting experimental decay curves can remove or obscure valuable information regarding the true noise spectrum.

\section{Proposed methodology}
\begin{figure*}[ht]
    \centering
    \includegraphics[width=\linewidth]{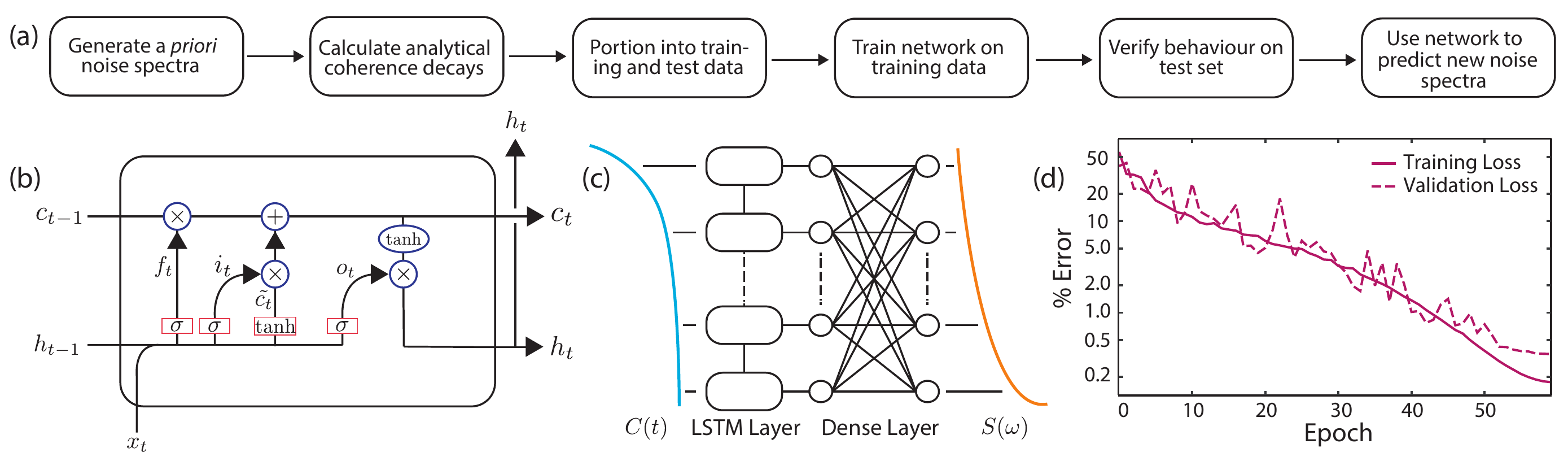}
    \caption{Our neural network approach. (a) Flow chart of the methodology. (b) A detailed view of a Long Short-Term Memory (LSTM) cell. $c_t$ is the cell state vector (providing the memory of the network), $x_t$ the cell input (one point in a coherence decay) and $h_t$ the cell output which is fed both to the output dense layer and the next cell in the series. Red boxes represent dense layers with either sigmoid ($\sigma$) or tanh activations. Blue circles represent pointwise operations. (c) A cartoon of the whole network: points on the coherence decay are fed into the LSTM layer, the output of the LSTM layer is then input into a dense layer and the ouptut of this dense layer is compared with a target noise spectrum. (d) An example network training graph showing training loss and validation loss (mean absolute percentage error) over training.}
    \label{fig:fig_2}
\end{figure*}

As illustrated in the discussion above, solving Eq.~\eqref{eq:2} to extract the noise spectrum specific to the qubit under investigation is non-trivial, however, the reverse operation (evaluating the coherence decay for a given noise spectrum and decoupling sequence) is relatively simple. This asymmetry lends itself to a neural networks-based learning approach to achieve a significant improvement in accuracy, owing to the capacity of neural networks to act as universal function approximators~\cite{Hornik1993, Huang2006}. 
To successfully implement the deep learning technique we first developed an efficient method for generating a sufficiently large and diverse set of training data. We split this training data into train/validation/test tranches and explored a selection of possible network architectures before selecting the most effective option and training it for optimal performance on the validation set~\cite{Goodfellow-et-al-2016}. Finally, we verified the chosen network's performance on the test set. A concise flow chart of the methodology is shown in Fig.~\ref{fig:fig_2}(a) with detailed description provided below.

\subsection{Generation of training data}
We simulated a variety of noise spectra assuming commonly applicable models described above and computed the corresponding coherence decay curves using Eq.~\eqref{eq:2}. Input and output data used in training have 151 data points each with log spacing in the time domain. Frequency range is determined by the time range of the coherence decay curves following the aforementioned relationship - $\omega = \tfrac{n\pi}{t}$ (as given by Eq.~\eqref{eq:1}). In the first instance, we generated three different forms of noise spectra for training purposes:

\begin{enumerate}
    \item A noise spectrum generated from a stretched exponential coherence decay. The delta function approximation shown in Eq.~\eqref{eq:3} is used to first produce an approximate noise spectrum for a given coherence decay. Eq.~\eqref{eq:2} is then used to produce a new coherence decay with an exact relationship to the noise spectrum. It is this new coherence decay that is then used as an input for neural network training.    
    \item ${1}/{f}$ noise, with functional form ${A}/{f^\alpha}$, generated using simulated parameters chosen to be in line with expected physical constants.
    \item A noise spectrum with a Lorentzian form:
    \begin{equation}
        \frac{\Delta^2\tau_c}{\pi}\frac{1}{1+(\omega\tau_c)^2}
        \label{eq:lorentz}
    \end{equation}
    
    where $\Delta$ and $\tau_c$ are coupling strength and correlation time, respectively, again generated using appropriate simulated parameters.
\end{enumerate}

In theory, these data can then be used to train a neural network to output the noise spectrum corresponding to a given coherence decay measured using a Hahn echo sequence. Specifically,  network training can be performed using coherence decays as inputs to the network, with noise spectra as the target outputs. 
Before training, the generated data must be split into training, validation, and test data sets. Training data is actively used to update the network parameters, whilst the network performance on validation data is monitored during training to avoid overfitting (see Fig.~\ref{fig:fig_2}(d))~\cite{Goodfellow-et-al-2016} and used for tuning the network hyperparameters (its overall structure, loss function etc.). The test data is held out for final evaluation of performance of the network on hitherto unseen data; error and loss rates quoted throughout this paper are always on the test set unless otherwise specified.

\subsection{Choice of Network Architecture}
Many different variants of neural network exist, the simplest being a deep feed forward neural network~\cite{Goodfellow-et-al-2016}, consisting of multiple layers each with a number of units. The first layer takes as input the training data $X$, with the outputs of each layer fed as input to the next. The final layer output is compared with the training data $y$ to calculate the network loss.
Each layer in a feed forward network can be represented by multiplication by a matrix of weights, $W$ followed by addition of a bias vector, $b$ and finally an activation function, $g$. The action of the layer, $h$, on its inputs $x$ can be written:

\begin{equation}
    h(x) = g(x^{\top}W + b)
\end{equation}

For each training step, the parameters of the neural network, $\phi$ are updated so that the overall function of the network $f$ is brought closer to the desired function $f^*$. This is done by taking the gradient of the chosen loss function with respect to the network parameters and updating those to reduce the loss, through a process known as back-propagation~\cite{Goodfellow-et-al-2016}. 

One problem with a feed forward network is that all inputs are connected to all units in each layer, which has the effect of destroying local correlations. For our purposes, a network architecture that preserves these local correlations is desirable. There are two obvious choices for working with time ordered data such as a coherence decay curve:  \emph{convolutional} and \emph{recurrent} (RNN) networks~\cite{Fukushima1980, Pearlmutter1989}.  After exploration the architecture found to be most effective for the task at hand was a Long Short-Term Memory network (LSTM)~\cite{Hochreiter1997}, a special case of the RNN. LSTMs introduce the capacity for long range correlations in input data to be preserved and several recent studies have found the LSTM paradigm to be effective in working with time series data~\cite{Arsene2019, Chen2018, Antczak2018ab}.

A simple RNN is a sequence of discrete units, each of which processes a single time step of the input data, $x_t$ and produces an output, $h_t$. This output is concatenated with the input of the next layer ($x_{t+1}$) to allow information earlier in the sequence to influence the output of layers later in the sequence. Although useful for maintaining short term correlations, one subtlety of the back propagation method of training neural networks is that early layers have relatively little impact on the gradient of the loss function of the network as a whole, meaning that early information is effectively `forgotten' by the network. To combat this, LSTMs introduce the concept of a `cell state', $c_t$, to which information can be added (remembered, $i_t$) or subtracted (forgotten, $f_t$), allowing the maintenance of important long term information. A schematic of a single LSTM cell can be seen in Fig.~\ref{fig:fig_2}(b) and its output is given by the following equations:

\begin{align}
    f_t &= \sigma_g(W_f(x_t \oplus h_{t-1})+ b_f)\\
    i_t &= \sigma_g(W_i(x_t \oplus h_{t-1}) + b_i)\\
    o_t &= \sigma_g(W_o(x_t \oplus h_{t-1}) + b_o)\\
    \Tilde{c_t} &= \tanh{(W_c(x_t \oplus h_{t-1}) + b_c)}\\
    c_t &= f_t\cdot c_{t-1} + i_t\cdot \Tilde{c_t}\\
    h_t &= o_t\cdot \tanh{(c_t)}
\end{align}

here $\sigma_g$ refers to the sigmoid activation function, $W$s refer to the weight matrices associated with each LSTM cell operation, $b$s refer to the bias vectors of each LSTM cell operation and $x_t$ and $h_t$ refer to the input and output vectors at the relevant time step. The $\oplus$ operation specifies a concatenation of the two vectors $x_t$ and $h_{t-1}$.
Many descriptions of the mathematical details and intuitions behind RNNs and LSTMs can be found~\cite{Goodfellow-et-al-2016, colah2015, Samek2019}. To produce an output of the correct size, the output from the LSTM is connected to a dense layer matching the size of the $y$ training data (see Fig.\ref{fig:fig_2}(c)). Although the activation function for the output layer of a neural network used for regression is typically a linear function, in this case we select an exponential function. This aids the training of the neural network as weights and biases are typically initialized for relatively small output values ($<5$) whilst the noise spectra used for regression in this case tend to have values $>10^4$. 

\begin{figure*}[ht]
    \centering
    \includegraphics[width=\linewidth]{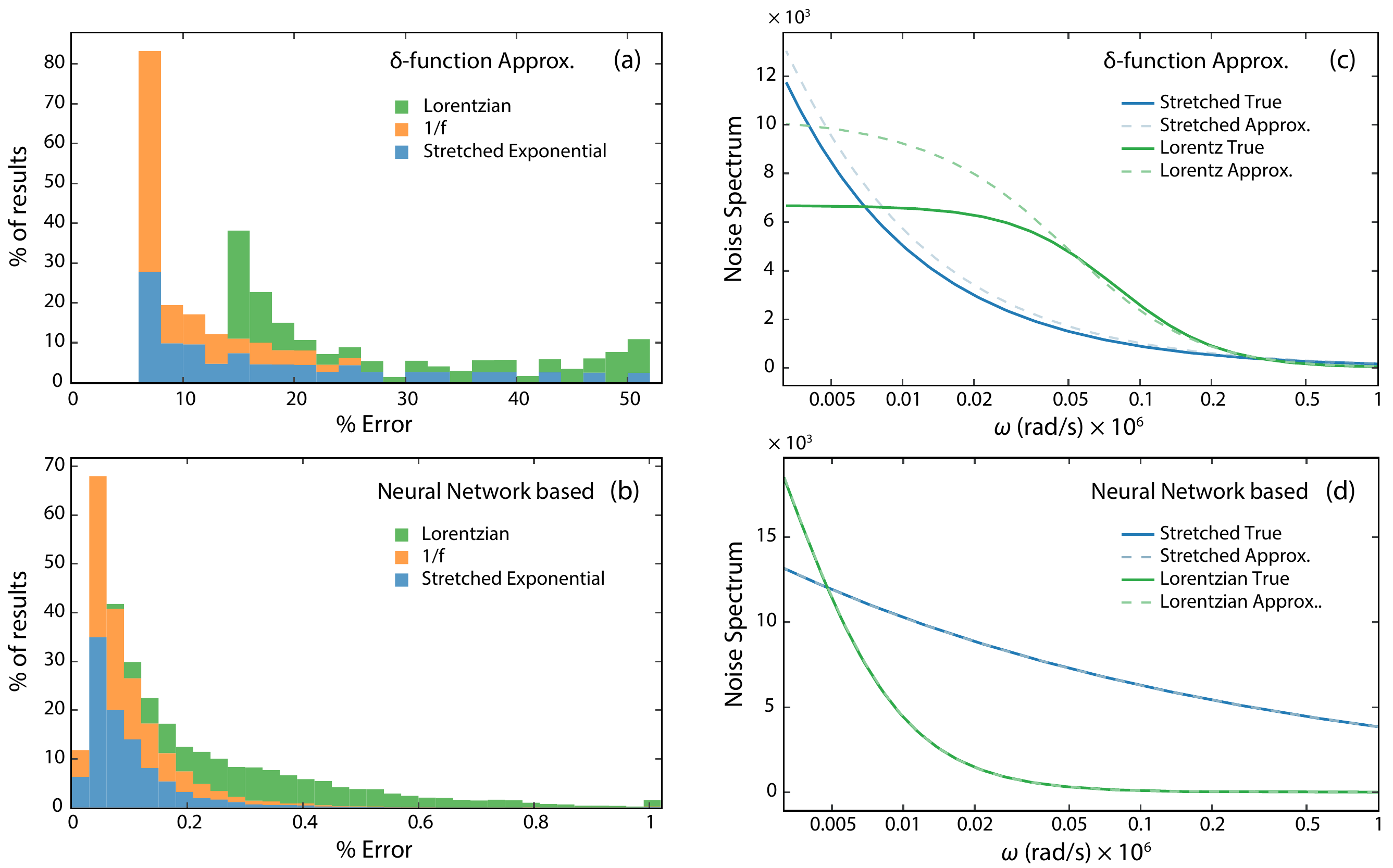}
    \caption{Comparison of two approaches for producing a noise spectrum from a measured coherence decay curve. 
    Stacked histograms of errors in the predicted noise spectra from the holdout test data set are shown using (a) a $\delta$ function approximation, and (b) our neural network. Three different forms of noise spectrum are studied, each with $\sim8,500$ individual noise spectra and associated decay curves. In (b), the tail of the histogram is clipped for clarity and the maximum error was 2.3\%. Also shown are the true (solid curves) and predicted (dashed curves) noise spectra obtained under (c) the $\delta$ function approximation and (d) the neural network approach. The noise spectra used are at the 50th percentile error for each approach.  
    }
    \label{fig:fig_3}
\end{figure*}

\subsection{Network Training}

The neural networks used for this study were defined using the Keras module of the Tensorflow 2.0 deep learning library distributed and maintained by Google~\cite{kerasab, tensorflow2015-whitepaper}. We carried out training using the previously discussed train/validation/test split. We use one cycle learning, as proposed by L. Smith~\cite{Smith2018ab}, to control learning rate during training. Hyperparameter tuning is undertaken to determine optimum network features using Bayesian search and the Weights and Biases library~\cite{wandb}.

\section{Results and Discussion}

Figure \ref{fig:fig_3} (and Fig. \ref{fig:supp_1}) compares the performance of the traditional $\delta$-function approximation for calculating the noise spectrum against our neural network-based approach. We show results for synthetic coherence decay curves generated from three different sources of noise: Lorentzian, ${1}/{f}$ and stretched exponential. The recorded performance of the neural network is for a \emph{single} network trained on the three noise sources simultaneously, applied to the three different test data sets.
Detailed statistics on the performance of the network are shown in Table \ref{table:1}. 

\begin{table}
\begin{tabular}{p{2.2cm}|c c|c c}
    & \multicolumn{2}{p{3cm}|}{\centering NN approach} & \multicolumn{2}{p{3cm}}{\centering$\delta$-function approx.} \\
    Noise model & Mean & $\sigma$ & Mean & $\sigma$ \\
    \hline
     \emph{Stretched Exp.} & 0.098 & 0.1 & 17.6 & 12.2 \\
     \emph{$\frac{1}{f}$} & 0.10 & 0.1 & 10.2 & 5.0 \\
     \emph{Lorentzian} & 0.39 & 0.21 & 28.4 & 13.7 \\
\end{tabular}
\caption{Comparison of error statistics in $\%$ in noise spectrum extraction using the neural network approach introduced here versus the $\delta$-function approximation for the pulse sequence filter function.}
\label{table:1}
\end{table}

\begin{figure*}[ht]
    \centering
    \includegraphics[width=\linewidth]{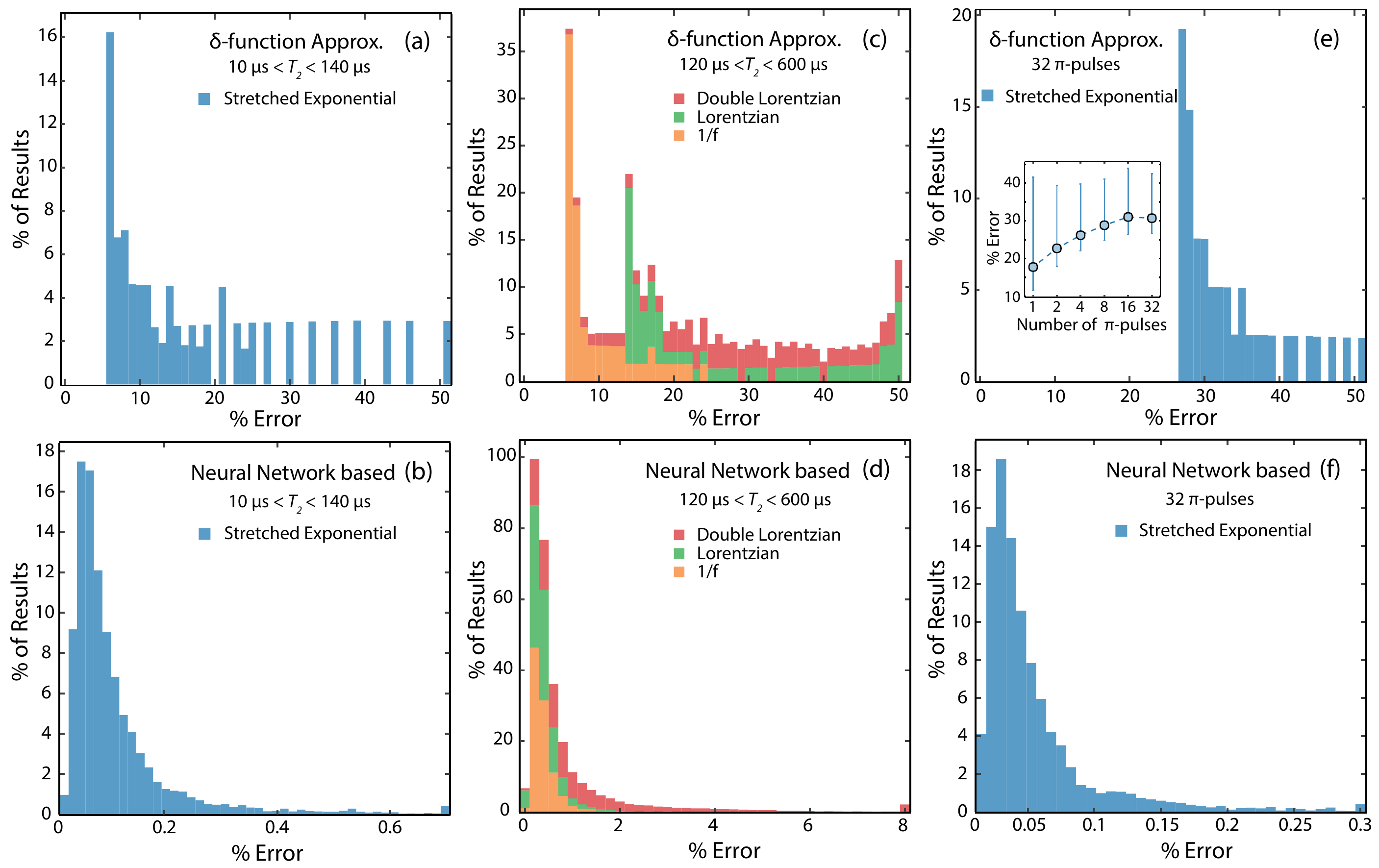}
    \caption{Comparison between estimated error in noise spectra predictions for three different neural networks and those obtained using a $\delta$-function approximation. Histograms of error rates for predicting noise spectra generated using Eq.\eqref{eq:3} from coherence decays with a coherence decay time constant $T_2$ between 10--140 $\mu$s obtained by means of (a) the $\delta$-function approach and (b) a trained neural network. Stacked histograms of error rates for predicting noise spectra having three different functional forms: $1/f$, Lorentzian, and double Lorentzian obtained by means of (c) the $\delta$-function approach and (d) a third trained neural network, the tail of the neural network distribution is clipped for clarity with maximum error in this case 200.1\% with high errors caused by noise spectra with large portions very close to 0 where small deviations in predicted noise spectrum magnify mean percentage error. Histograms of error rates for predicting stretched exponential noise spectra generated from coherence decays constructed by implementing dynamical decoupling protocol comprising 32 $\pi$-pulses extracted by means of (e) the $\delta$-function approach and (f) a second trained neural network; inset in (e) shows mean percent errors along with the maximum deviation as a function of number of $\pi$-pulses for conventional technique.}
    \label{fig:4}
\end{figure*}

Our results show that the neural network approach significantly outperforms the methodology based on a $\delta$-function approximation to the sequence filter function for deducing noise spectra from qubit coherence decay curves. Of particular interest is that a single network is able to successfully reproduce noise spectra of different functional forms. A current limitation of this approach is that the network has only been trained on a set of coherence decays with a decoherence time constant $T_2$s in a relatively small range between approximately 120 $\mu$s to 600 $\mu$s. In principle, this can be circumvented by training a network on the same principles but for different length coherence decays. For example, in Fig.~\ref{fig:4}(a) and (b) a network is trained on coherence decays with $T_2$s between 10--140 $\mu$s, generated using the phenomenological stretched exponential method discussed above. This neural network approach can therefore be applied generally, given additional training and data generation to cover the parameter range of interest. A more generalised approach would use the time vector data for training, in addition to the signal data, and represents a promising direction for further development.

\begin{table}

\begin{tabular}{p{2.7cm}|c c|c c}
    & \multicolumn{2}{p{2.8cm}|}{\centering NN approach} & \multicolumn{2}{p{2.8cm}}{\centering$\delta$-function approx.} \\
    Noise Model & Mean & $\sigma$ & Mean & $\sigma$ \\
    \hline
     \emph{$\frac{1}{f}$} & 0.41 & 0.46 & 10.2 & 5.0 \\
     \emph{Lorentzian} & 0.42 & 0.40 & 28.4 & 13.7 \\
     \emph{Double Lorentzian} & 1.72 & 3.63 & 30.8 & 11.9 \\
\end{tabular}
\caption{Error statistics in $\%$ for neural network and $\delta$ function noise spectrum approach, for a network trained on ${1}/{f}$, Lorentzian and double Lorentzian noise spectra}
\label{table:2}
\end{table}

\begin{figure*}[ht]
    \centering
    \includegraphics[width=\linewidth]{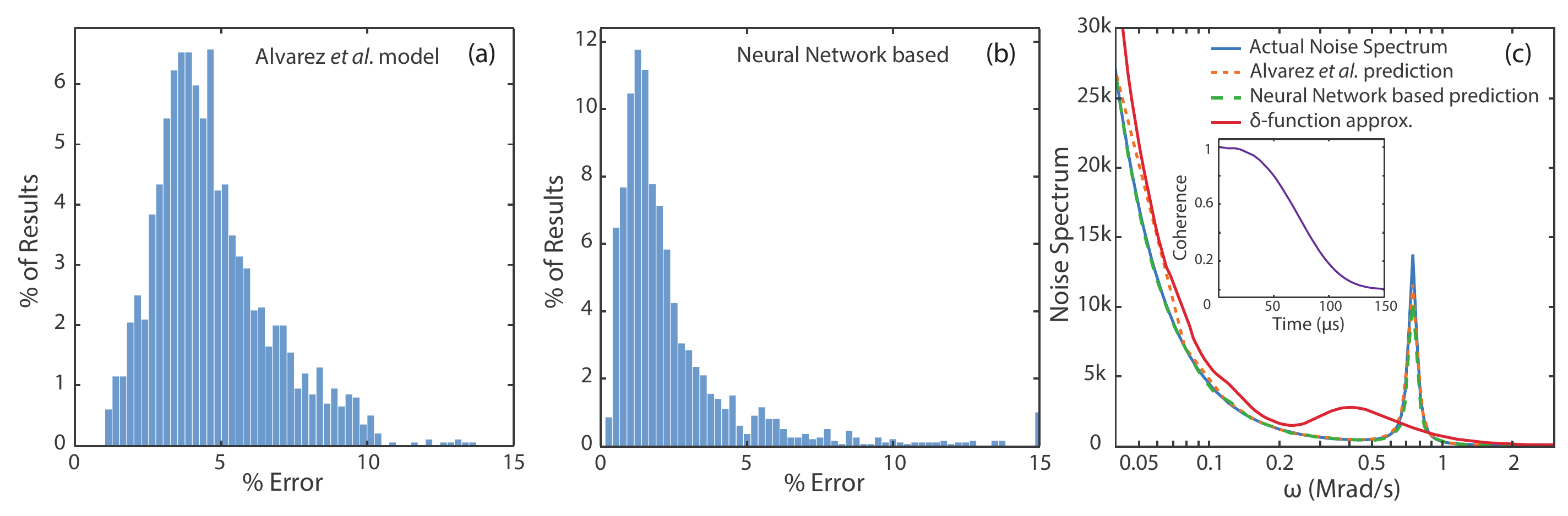}
    \caption{Comparison of neural network approach with an established approach. a) Histograms of errors in noise extraction when employing Alvarez \textit{et al.} methodology~\cite{Alvarez2011}, b) histograms of errors in predicting noise via trained neural network, c) A plot comparing predictions of different methodologies for a representative noise spectrum. The inset shows the decoherence curve corresponding to the underlying noise spectrum. Note: All the artificial noise spectra were generating using a $1/f$ model along with a Lorentzian-shaped feature whose parameters are chosen randomly.}
    \label{fig:4_AS}
\end{figure*}


Next, we demonstrate the performance of neural network based approach for quantum systems experiencing multiple simultaneous noise sources. Figs.~\ref{fig:4}(c) and \ref{fig:4}(d) show the results of a network that is trained to predict noise spectra arising from three different models: ${1}/{f}$, Lorentzian and double Lorentzian (sum of two Lorentzians with different $\Delta$, and $\tau_c$ following Eq.~\eqref{eq:lorentz}). Double Lorentzian noise spectra are characteristic of quantum systems experiencing noise from two uncorrelated sources, as observed in near-surface spins such as NV centers in diamond~\cite{Romach2015}. Our results show how a single trained network can successfully predict cases where a coherence decay is a result of a system experiencing multiple noise sources, whilst also differentiating multiple single noise sources. Furthermore, training data comprising varied models allows the network to learn complex correlations and subsequently output robust predictions in the context of previously unseen data (see Appendix B, Fig. \ref{fig:supp_2}). This proof-of-principle demonstration suggests that expanding the range of noise spectra that the network is able to identify, particularly where a noise spectrum is made up of multiple different noise sources, is a fruitful direction for future research.

Having shown the robustness of the neural network approach to a range of coherence decay times and a combination of noise spectra, in Fig.~\ref{fig:4}(e) and (f) we show the performance of a network trained to reproduce noise spectra from coherence decays generated using the phenomenological method described above but in this case using a filter function associated with a 32 $\pi$ pulse CPMG sequence. The similarity between the $\delta$-function and the actual filter function of a CPMG sequence with sufficiently large number of $\pi$ pulses is generally used to justify the approach of employing Eq.~\eqref{eq:3} to extract noise spectra. Surprisingly, we find the $\delta$-function approach to be less accurate in this case than when a Hahn echo filter function is used as shown in Fig.~\ref{fig:4}. The neural network appears unaffected by the change in pulse sequence and remains able to successfully reproduce the noise spectra with a low error rate. We speculate that the reason for the poor performance of the $\delta$-function approach in this case is that, whilst the central peak of the filter function becomes narrower at higher pulse numbers, the contributions of higher frequency harmonics become more significant~\cite{Loretz2015}(also see Fig.~\ref{fig:Fig1}(e) and (f)). Therefore, for monotonic noise spectra, such as those considered in Fig.~\ref{fig:4}, the advantages of using multiple-$\pi$-pulse measurements together with $\delta$-function approximation are not evident. Nevertheless, the implementation of multi-$\pi$-pulse protocols becomes indispensable for non-monotonic feature extraction owing to the narrowness of the filter function that they produce. Hence, researchers have developed elaborate strategies for mitigating the errors introduced by the higher harmonics of the CPMG filter function.

One of these state-of-the-art approaches, developed by Alvarez and Suter~\cite{Alvarez2011}, employs sequences involving multiple $\pi$-pulses to infer the underlying noise spectrum. The approach relies on the fact that if the sequence maintains a constant $\tau$, and measures the qubit decay as a function of number of $\pi$-pulses then the decay rate, $R$, of the qubit during dynamical decoupling becomes time independent and its relationship with the noise spectrum can be defined as follows:

\begin{equation}
\label{eq:13}
    R = \sum_{k=1}^{\infty} A_k^2 S(k\omega_0)
\end{equation}

Where $\omega_0=\tfrac{\pi}{\tau}$ and $A_k$s are related to the sensitivity amplitudes of the filter function at the probing frequencies (see Appendix C, Fig \ref{fig:supp_3}(a)). Therefore, by measuring signal at carefully selected values of $\tau$ and repeating the experiment several times by employing multiple $\pi$-pulse decoupling sequences, accurate estimation of bath noise spectrum can be achieved (see Fig \ref{fig:supp_3} for more details on the actual methodology). 

\begin{figure*}[ht]
    \centering
    \includegraphics[width=\linewidth]{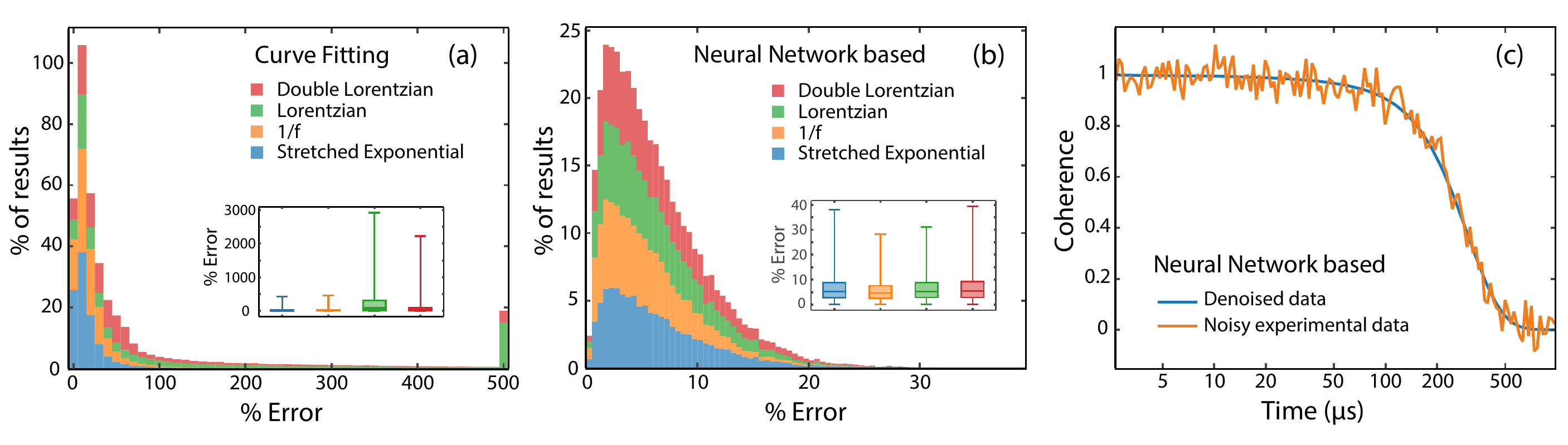}
    \caption{Denoising the experimental data using curve fitting or neural networks. Stacked histograms of errors in smoothing synthetic experimental decoherence curves from four classes of noise model are shown using two denoising approaches: a) curve fitting to Eq.\eqref{eq:4}, or b) the neural network approach. In both cases, insets show mean errors and the maximum deviation. The percent errors are calculated using logarithmic values of decoherence curve.
    c) A synthetic noisy decoherence curve (from a $1/f$ noise model) along with a denoised curve obtained from the neural network.}
    \label{fig:4_denoising}
\end{figure*}

To benchmark the neural network based approach against such more elaborate experimental methods, we have trained a network with a set of noise spectra that have an underlying $\frac{1}{f}$ functional form overlaid with an additional Lorentzian-shaped feature with randomly selected amplitude, peak position, and width. Fig.~\ref{fig:4_AS} demonstrates that the neural network-based methodology performs at least as well as Alvarez and Suter, despite the former using only a simple Hahn-echo experiment. Both approaches are significantly superior to the simple $\delta$-function approach. Notably, the results establish that the network is capable of extracting more complex non-monotonic noise spectra with high accuracy, circumventing the need of lengthy and technically challenging measurements. Further work will be required to upgrade the network, both in terms of the architecture and the diversification of the training data, for extending its applicability to varied qubit environments. 

There exist more intricate strategies, for example, discrete prolate spheroidal sequences (DPSSs)~\cite{Norris2018} have a filter function that is devoid of additional lobes and higher harmonics resulting in suppression of spectral leakage, while a Gaussian enveloped dynamic sensitivity control sequence (gDYSCO)~\cite{Romach2019} allows generation of a filter function with minimal spectral leakage at the expense of reduced sensitivity and broader main peak. 
Though more effective than the basic CPMG approach, these sequences can be challenging to implement in practice and require repeated measurements~\cite{Norris2018,Romach2019}, while in contrast our neural network-based approach has been shown to accurately extract underlying noise spectra using only a single measurement of a Hahn echo decay curve.  Nevertheless, in principle, elements of both techniques could be combined in the pursuit of even better noise spectrum extraction, while minimising the increase in experimental complexity.

\begin{table}

\begin{tabular}{p{2.7cm}|c c|c c}
    & \multicolumn{2}{p{2.8cm}|}{\centering NN approach} & \multicolumn{2}{p{2.8cm}}{\centering curve fitting} \\
    Noise Model & Mean & $\sigma$ & Mean & $\sigma$ \\
    \hline
     \emph{Stretched exp.} & 6.49 & 4.84 & 16.79 & 20.53 \\
     \emph{$\frac{1}{f}$} & 5.52 & 3.82 & 21.21 & 23.06 \\
     \emph{Lorentzian} & 6.35 & 4.42 & 233.67 & 341.21 \\
     \emph{Double Lorentzian} & 6.72 & 4.91 & 96.72 & 156.02 \\
\end{tabular}
\caption{Error statistics in $\%$ of $\log C(t)$ for neural network and stretched exponential fitting denoising approaches.}
\label{table:denoising}
\end{table}

\begin{figure*}[ht]
    \centering
    \includegraphics[width=\linewidth]{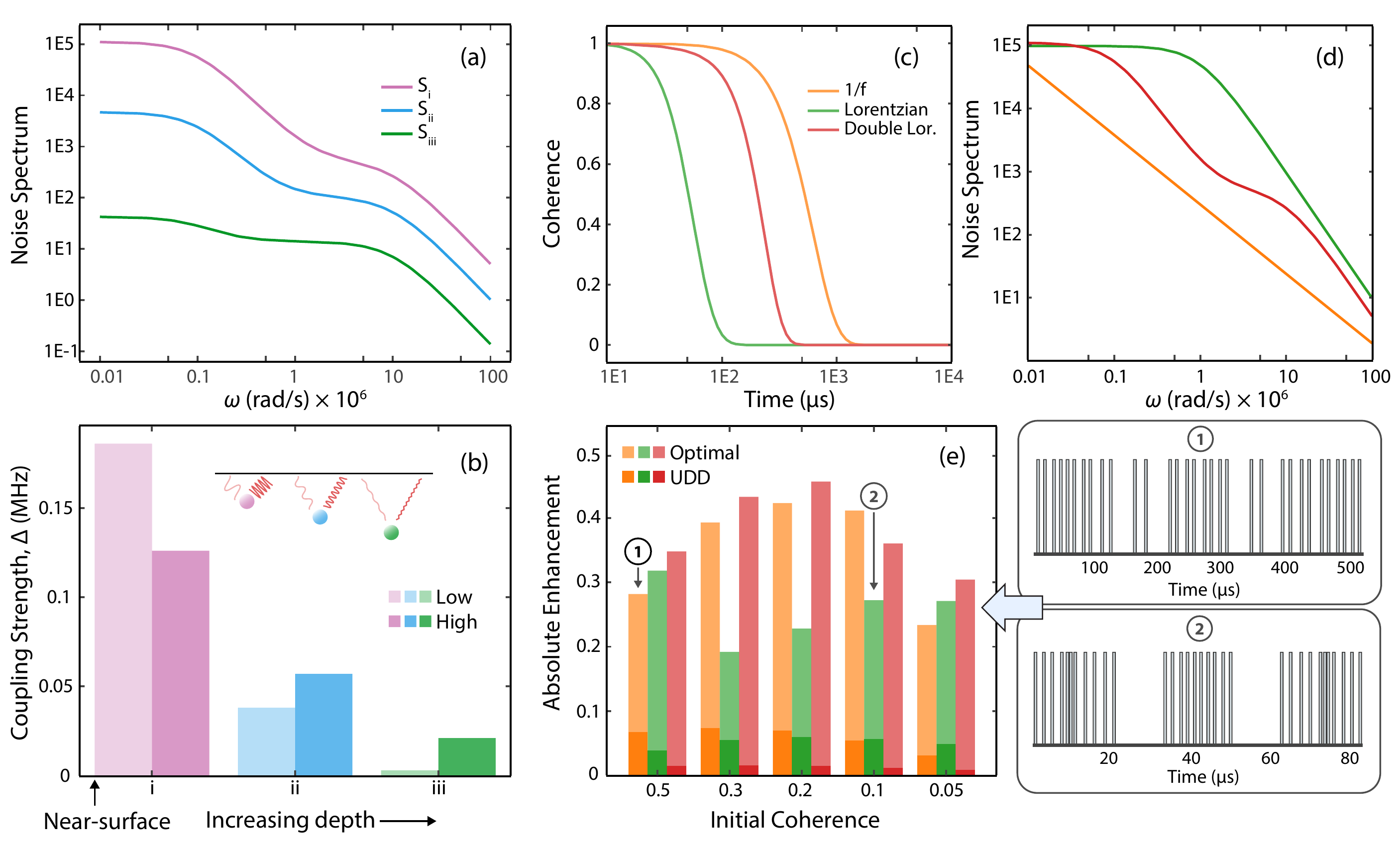}
    \caption{Potential applications of the methodology include defect depth prediction~\cite{Romach2015} and the generation of customised dynamical decoupling sequences. (a) Simulated double Lorentzian-type noise spectra characteristic of near-surface spin-defects, for different increasing depths (i--iii). (b) Defect-bath coupling strengths for low and high frequency noise corresponding to noise spectra shown in (a); the variations in the coupling strengths are indicative of changes in the defect position with respect to the surface. Generation of optimized dynamical decoupling sequences: (c) Specimen decoherence curves derived by applying 32 $\pi$-pulse CPMG to the noise spectra shown in (d). (e) Histograms demonstrating coherence improvement after implementing the SLSQP algorithm to optimize $\pi$-pulse positions; enhancements obtained by employing UDD protocol~\cite{Uhrig2007} are also included for comparison. Extracted optimal protocols corresponding to two histograms denoted by \raisebox{.5pt}{\textcircled{\raisebox{-.1pt} {\tiny{1}}}} and \raisebox{.5pt}{\textcircled{\raisebox{-.1pt} {\tiny{2}}}} are shown alongside the bar plot, illustrating an increasing divergence from the equivalent CPMG sequence as coherence decreases. Initial coherence refers to the value of coherence for a 32-pulse CPMG sequence, absolute enhancement refers to the increased coherence at the same point in time achieved with a UDD and optimized sequence. If the qubit had an initial coherence of 0.5 at 100~$\mu$s with a CMPG-32 sequence and had a coherence of 0.8 at 100~$\mu$s with an optimized sequence, then the absolute enhancement is 0.3.}
    \label{fig:5}
\end{figure*}

\subsection{Handling Experimental Data}

The utility of our proposed approach requires a robust method for ingesting experimental coherence curves, including unavoidable experimental measurement noise. A traditional approach would be to apply a least squares fit of a stretched exponential function (given in Eq.~\ref{eq:4}) to the noisy data to gain a best approximation to the underlying decay. However, such an approach necessarily makes an assumption regarding the particular form of the noise, restricting the possible noise spectra that can be inferred to those that produce a stretched exponential coherence decay. 

We propose instead a function-agnostic approach to removing experimental noise. Once again, we look to LSTM based neural networks to perform this task --- an approach that has already been successfully used for denoising of electrocardiogram (ECG) data~\cite{Antczak2018ab}. We trained the network by applying random noise of $\pm 5\%$ to the coherence curves generated from a) the three different noise models described above, or b) the phenomenological stretched exponential curves. This was used as the input $X$ data, whilst the original noiseless curves were used as the target $y$. To evaluate the success of this neural network technique we compare its ability to reconstruct a given coherence decay against the approach of fitting a stretched exponential to the same noisy data. 

To infer the impact of the errors on reconstruction of the noise spectrum, we compare the mean absolute percentage errors in $\log C(t)$ of ground truth versus the noisy data, as the noise spectrum depends explicitly on this as seen in \ref{eq:2} --- the results are shown in Fig.~\ref{fig:4_denoising}, with detailed statistics in Table~\ref{table:denoising}. Performance for the neural network is comparable to that of fitting for stretched exponential and ${1}/{f}$ coherence decays, although it does show some improvement. However, the advantage of the network becomes clear when we examine errors for coherence decays derived from Lorentzian and double Lorentzian noise spectra: whilst the neural network performs similarly on all decay types, the error in curve-fitting increases drastically for these cases. The results demonstrate that a functional-form agnostic approach to denoising --- as permitted by the neural network --- is essential for an accurate and general extraction of the noise spectrum from experimental data.

\section{Outlook}

Accurately deducing the noise environment of a quantum system has many potential applications: for qubits, knowledge of the noise spectrum can be used to extend coherence times through bespoke dynamical decoupling sequences~\cite{Biercuk2009, Cywinski2008}, while for spin defects in solid state systems, the noise spectrum has been shown to provide information on parameters such as defect depth~\cite{Romach2015}. Figure~\ref{fig:5} illustrates both such applications. 

Using an approach based on the $\delta$-function approximation,  Romach \emph{et al.} find evidence for a double Lorentzian form for $S(\omega)$ for shallow NV centers in diamond, attributed to the presence of two distinct noise sources with differing correlation times. The faster correlation time was attributed to surface-modified phonons and the slower to spin-spin coupling between a bath of surface spins. The coupling strength, $\Delta$, to each of these noise sources decreases as defect depth increases. Our technique provides a potential method for accurately estimating defect depth through simple coherence decay measurements, at a much lower experimental cost than previous approaches. Although, the presented noise spectra in Figure~\ref{fig:5} (a) are artificial, the objective here is to show that the precise extraction of parameters corresponding to the underlying noise spectrum such as coupling strengths (in line with Eq.~\ref{eq:lorentz}) provides an excellent means to detect qubit's surface proximity as sketched in Figure~\ref{fig:5} (b).

Figures~\ref{fig:5}(c)-(e) show the application of dynamical decoupling optimization to three different classes of noise spectra. Using the sequential least squares programming (SLSQP) minimization technique~\cite{Kraft1994}, provided by the SciPy library~\cite{2020SciPy-NMeth}, we significantly enhance residual coherence over what can be achieved using a 32-pulse CPMG or UDD sequence. The SLSQP algorithm allows for the constrained optimisation of $\pi$-pulse position, starting from 32-pulse CPMG positions, for a given total sequence time to minimize the value of $\chi(t)$, which in turn maximizes the value of $C(t)$. The optimization process is constrained to ensure pulse ordering is preserved and to prevent overlaps between pulses. Underlying noise spectra are used to synthesise coherence decay curves under 32-pulse CPMG, and to generate bespoke pulse sequences (two examples of which are illustrated) that enhance the coherence at specific points in time. While UDD offers some limited enhancement in coherence over CPMG, substantial increases are seen using the optimized sequences, up to values of 4--8$\times$. 

Whilst the results presented here are promising, the potential avenues for improvement are equally clear. The technique currently requires a specific input dimension and time scale, though an encoder-decoder architecture, as seen in sequence-to-sequence models and transformers, provides a possible solution~\cite{Vaswani2017, Wu2020ab}. In addition, whilst the optimization techniques described above appear successful, the application of deep learning techniques to optimising qubit coherence times is a promising route for exploration. Furthermore, recent results have shown the successful application of deep reinforcement learning to quantum sensing problems, suggesting they may well have application in this field~\cite{Schuff2020, fiderer2020ab}. It would also be instructive to explore the impact of pulse errors on the noise-spectrum coherence decay relationship. Pulse errors are unavoidable in experiments and being able to account for their effects would add further applicability to the techniques developed here. We anticipate that the performing experiments under the influence of artificial noise sources and known pulse errors will allow greater insight into these effects. In addition, there has been recent theoretical work to explore the generation of coherence decays from noise spectra when pulse errors are present~\cite{He2018}. Incorporating these techniques into the training data generation process detailed above would allow the training of networks where such pulse errors are accounted for. 

In summary, we have demonstrated a multifaceted toolbox that uses neural networks to accurately deduce the  environmental noise spectrum of a qubit. We show that an LSTM network, properly trained with diverse training data, is capable of predicting the precise noise spectrum from a coherence decay curve recorded using dynamical decoupling protocols comprising variable numbers of $\pi$ pulses. To treat experimental data with measurement noise, we again employ an LSTM network to perform effective denoising that preserves the original functional form of the decoherence curve. This unprejudiced way of smoothing noisy experimental data avoids the use of a predetermined fitting function, enabling more accurate reconstruction of the qubit environment. Using these deep learning techniques one can even extract information about multiple noise sources, which can then be used to infer salient properties of the qubit under investigation. Finally, we show how the extracted noise spectrum can be used to generate a customized dynamical decoupling sequence that enhances coherence time significantly beyond what can be achieved with standard protocols. We have surveyed a variety of possible avenues related to the application of deep learning for problems related to qubit decoherence and have established that approach has several potential applications in different quantum technologies. Despite this promise, these concepts remain in their infancy and there is much scope for further improvement and development. 

\section{Acknowledgments} 
We thank Prasanna Vaidya, Gyde, India, for the fruitful discussions on neural networks. This research has received funding from the Engineering and Physical Sciences Research Council (EPSRC) via the Centre for Doctoral Training in Delivering Quantum Technologies (EP/L015242/1), as well as from the European Research Council (ERC) via the LOQO-MOTIONS grant (H2020-EU.1.1., Grant No. 771493).

\section{Appendix A: Alternative comparison between $\delta$-function-based vs Neural network-driven approaches}

\begin{figure*}[h!]
	\centering
	\includegraphics[width=0.9\linewidth]{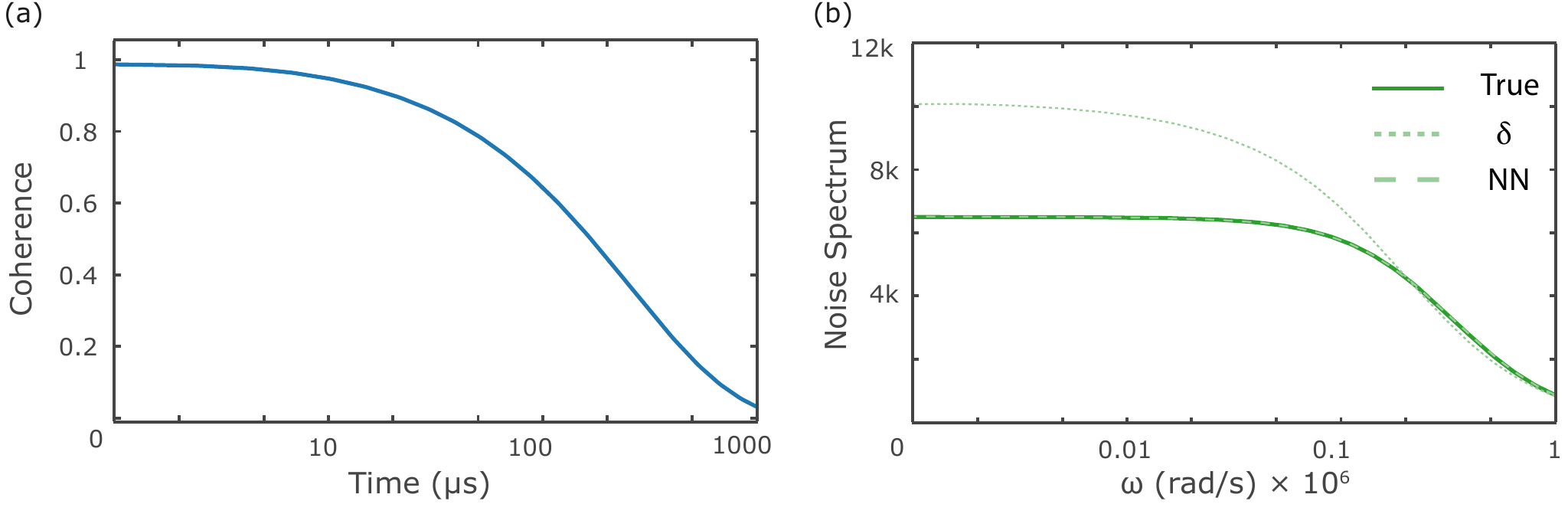}
	\caption{Figure showing neural network vs $\delta$-function approximation performance for inferring the Lorentzian noise spectrum shown in (b) from the coherence decay curve shown in (a). The true (target) noise spectrum is shown in solid green in (b), the $\delta$-function approximation result is shown by a dotted pale-green line, the neural network result is shown by a dashed pale-green line.}
	\label{fig:supp_1}
\end{figure*}

Fig. 3 in the main text compares the performance of the $\delta$ function approximation and the neural network approach at the 50th percentile in (c) and (d). For added transparency, we include here the performance of both approaches for the reconstruction of the same Lorentzian noise spectrum in figure \ref{fig:supp_1}

\section{Appendix B: Network performance on unseen noise forms}

To evaluate the generality of the neural networks used here and their potential for application to potentially novel qubit platforms with unknown noise forms, we take the neural network used to evaluate noise spectra of forms $1/f$, Lorentzian and double Lorentzian in the main text and apply it to coherence decays derived from noise spectra with $1/f$ + Lorentzian form. This is performed with no further training of the network. The results are shown below in figure \ref{fig:supp_2}. In addition, the figure shows the performance of the same network after being trained on a set of data with the target $1/f$ + Lorentzian form.

Notably the untrained neural network performs significantly better than the $\delta$-function approximation approach, with a mean error of 11.9\% vs 24.4\%. The neural network error distribution has a long tail due largely to noise spectra with significant portions valued close to 0, where mean percentage error is magnified. 

\begin{figure*}
	\centering
	\includegraphics[width=0.9\linewidth]{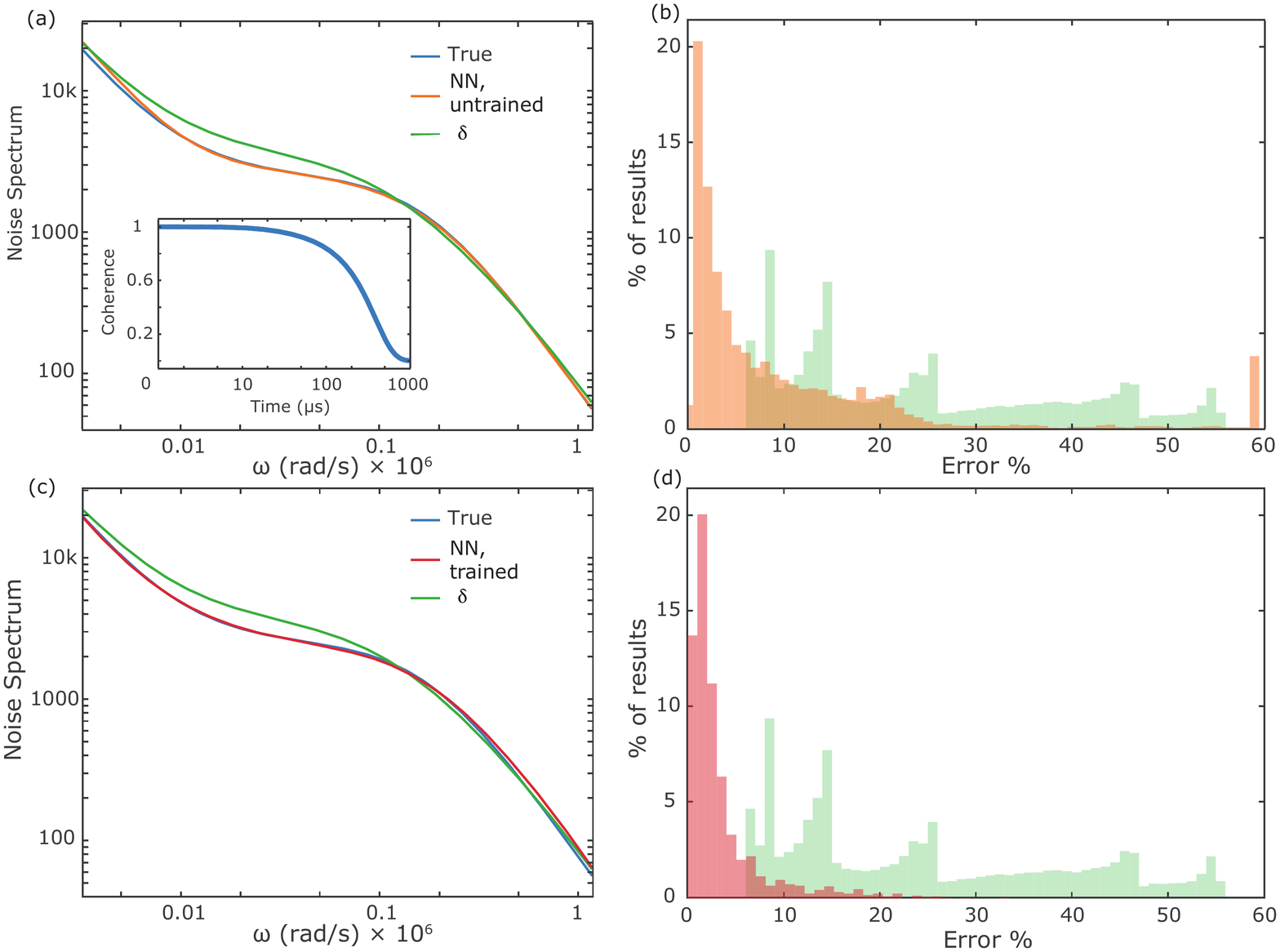}
	\caption{Figure showing neural network performance for inferring noise spectra from coherence decay curves derived from $1/f +$ Lorentzian noise, having been trained only on $1/f$, Lorentzian and double Lorentzian data. (a) shows an example target noise spectrum in blue, resulting in the coherence decay curve in the inset, with the approximate noise spectra derived from the $\delta$ function approximation and the untrained neural network. (b) shows the distribution of errors across a set of data of this functional form. The mean error for the untrained neural network is 11.9\% whilst the mean error for the $\delta$-function approximation is 24.4\%. The distribution of the untrained neural network has a long tail due to noise spectra with large portions close to 0, where mean percentage error is magnified. (c) and (d) show the same data, but with the neural network now fine-tuned on a small $1/f$ + Lorentzian dataset (entirely separate from the data upon which it is evaluated to maintain train/test split). In this case the mean error drops to 3.5\%.}
	\label{fig:supp_2}
\end{figure*}

To evaluate the network performance further, a small amount of data is generated and the network trained further (fine-tuned) and its performance evaluated once more. After this extra training, mean error for the network is reduced to 3.5\% and the tail of the distribution becomes significantly shorter. Generation of this extra data took approximately 10 minutes, whilst additional training required 15 minutes. This extra step is taken to highlight the adaptability and practicality of the approach described in the manuscript. If a new noise form is identified that the network is not trained to reconstruct, then only a small overhead is required to fine-tune the network for this task, with the expectation of greatly increased accuracy of noise-spectrum reconstruction.

\begin{figure*}[ht]
	\centering
	\includegraphics[width=0.95\linewidth]{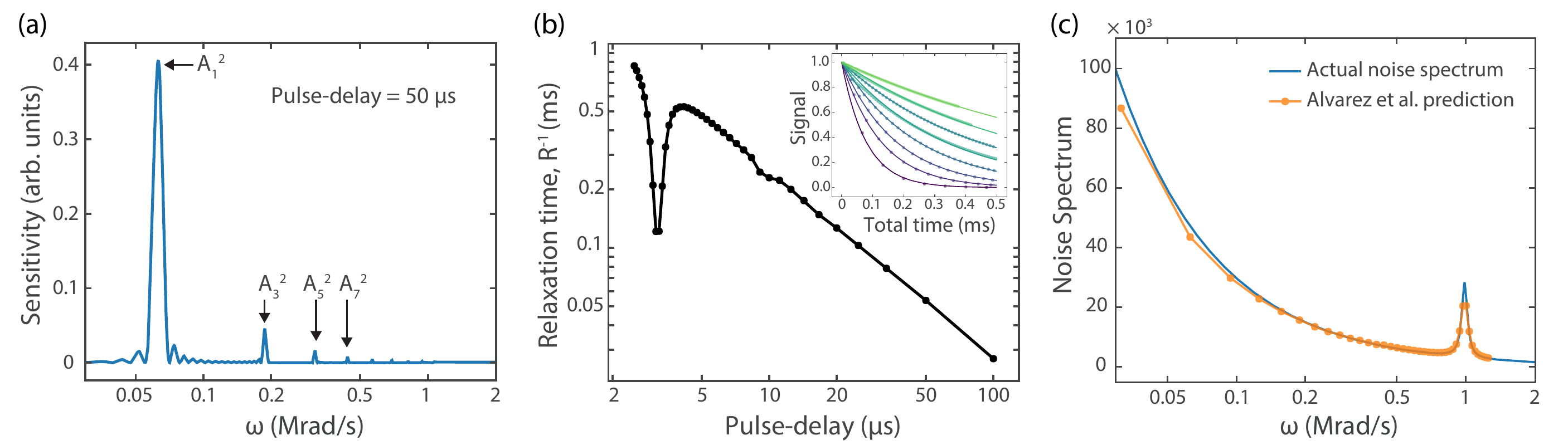}
	\caption{Figure illustrating application of Alvarez-Suter methodology to the simulated data to predict the underlying noise spectrum (a) Plot demonstrating how $A_k$s are associated with the sensitivities of the filter function corresponding to a given pulse-delay. (b) Relaxation times extracted via fitting the signal at various pulse-delays. The inset shows the actual signal for select pulse-delays. Each data-point here corresponds to the value of coherence recorded for the given pulse-delay and the total duration of the sequence; the total duration can be varied by changing the number of $\pi$-pulses present in the protocol. (c) Comparison between the actual noise spectrum and the predicted noise spectrum.}
	\label{fig:supp_3}
\end{figure*}

\section{Appendix C: Alvarez-Suter methodology}
We provide explicit illustration of the approach propose by Alvarez \textit{et al.}~\cite{Alvarez2011} in Fig \ref{fig:supp_3}. Note that an additional improvement in the noise spectrum prediction can be obtained if an approximate functional form of the noise spectrum can be derived. However, this approach is not readily generalizable and thus Eq. 13 in main manuscript has been used here for performing the calculations.

\clearpage
\bibliography{MyCollection} 


\end{document}